\documentclass[%
reprint,
superscriptaddress,
 amsmath,amssymb,
aps,
prb,
groupedaddress,
]{revtex4-1}
\usepackage{graphicx}
\usepackage{dcolumn}
\usepackage{bm}
\usepackage{colortbl}
\usepackage{hyperref}
\newcolumntype{R}[1]{>{\raggedleft\arraybackslash }b{#1}}
\newcolumntype{L}[1]{>{\raggedright\arraybackslash }b{#1}}
\newcolumntype{C}[1]{>{\centering\arraybackslash }b{#1}}

\begin{document}


\title{Thermal Control of the Magnon-Photon Coupling in a Notch Filter coupled to a Yttrium-Iron-Garnet/Platinum System}

\author{Vincent Castel}
\email{vincent.castel@imt-atlantique.fr}
\affiliation{Lab-STICC (UMR 6285), CNRS, IMT Atlantique, Technopole Brest Iroise, 29200 Brest, France
}%
\author{Rodolphe Jeunehomme}
\affiliation{IMT Atlantique, Technopole Brest Iroise, 29200 Brest, France
}%
\author{Jamal Ben Youssef}
\affiliation{Lab-STICC (UMR 6285), CNRS, Universit\'e de Bretagne Occidentale, 6 Avenue Victor le Gorgeu, 29200 Brest, France
}%
\author{Nicolas Vukadinovic}
\affiliation{Dassault Aviation, 78 quai Marcel Dassault, 92552 St-Cloud, France
}%
\author{Alexandre Manchec}
\affiliation{Elliptika (GTID), 29200 Brest, France
}%
\author{Fasil Kidane Dejene}
\affiliation{Max Planck Institute of Microstructure Physics, Weinberg 2, D-06120 Halle, Germany
}%
\author{Gerrit E. W. Bauer}
\affiliation{Institute for Materials Research, WPI-AIMR, and Center for Spintronics Research Network, Tohoku University, Sendai 980-8577, Japan}
\affiliation{Zernike Institute for Advanced Materials, University of Groningen, Nijenborgh 4, 9747 AG Groningen, The Netherlands}

\date{\today}

\begin{abstract}
We report thermal control of mode hybridization between the ferromagnetic resonance (FMR) and a planar resonator (notch filter) working at 4.74 GHz. The chosen magnetic material is a ferrimagnetic insulator (Yttrium Iron Garnet: YIG) covered by 6 nm of platinum (Pt). A current-induced heating method has been used in order to enhance the temperature of the YIG/Pt system. The device permits us to control the transmission spectra and the magnon-photon coupling strength at room temperature. These experimental findings reveal potentially applicable tunable microwave filtering function.

%
\end{abstract}


\maketitle

\section{Introduction}

Magnon-photon coupling has been investigated in microwave resonators (three-dimensional cavity \cite{Kang2008,Gollub2009,Zhang2014,Lambert2015,Haigh2015,Bai2015,Maier-Flaig2016,Bai2016,Harder2016,Harder2017} and planar configuration \cite{Stenning2013,Bhoi2014,Klingler2016}) loaded with a ferrimagnetic insulator such as Yttrium Iron Garnet (YIG, thin film and bulk). More recently, research groups \cite{Bai2015,Maier-Flaig2016,Bai2016} have developed an electrical method to detect magnons coupled with photons. This method has been established by placing a hybrid YIG/platinum (Pt) system into a microwave cavity, showing distinct features not seen in any previous spin pumping experiment but already predicted by Cao et al. \cite{Cao2014}. These later studies have been realized in a 3D cavity (with high $Q$ factor), but insertion of an hybrid stack including a highly electrical conductor such as platinum has been reported to induce a drastic enhancement of the intrinsic loss rate.

Here, we demonstrate thermal control of the magnon-photon coupling at room temperature in a compact, design based on a notch filter resonating at 4.74 GHz and a hybrid YIG/Pt system. The thermal control of frequencies hybridization at the resonant condition is realized by current-induced Joule heating in the Pt film, which causes an out-of-plane temperature gradient. The planar configuration of our rf device has a limited $Q$ factor but avoids the negative impact of the YIG/Pt stack.

\section{Compact design description}
\begin{figure}[h]
	\includegraphics[width=8cm]{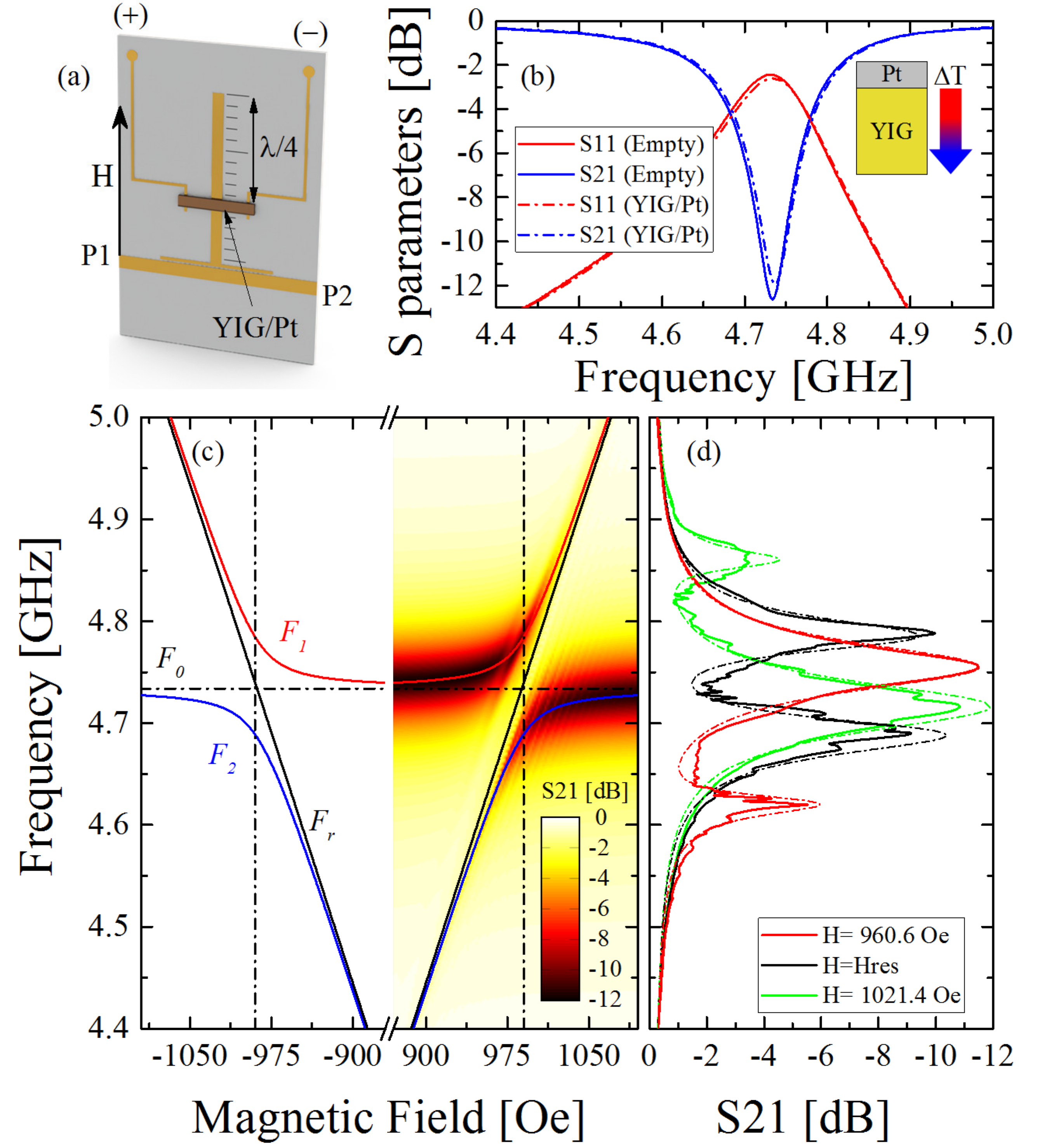}
	\caption{\label{fig:Fig1} 
		(a) Experimental setup: A vector network analyzer (VNA) is connected to a 50 $\Omega$ microstrip line which is capacitively coupled to the resonator. The hybrid YIG/Pt system is placed at 0.25 $\lambda$. (+) and (-) are the electrical contacts which permit to detect (inject) a dc charge current from (at) the edges of the Pt layer. (b) Frequency dependence of $S$ parameters with and without the YIG/Pt system. (c) Magnetic field dependence of the frequency: Observation of the strong coupling regime via the anti-crossing fingerprint. The color map illustrates the magnitude of $S21$ from 0 to -12 dB. (d) Frequency dependence of $S21$ at for different applied magnetic fields using the analytic solution from Ref. \cite{Harder2016}.
	} 
\end{figure}
The used insulating material consists of a single-crystal (111) Y$_3$Fe$_5$O$_{12}$ (YIG) film grown on a (111) Gd$_3$Ga$_5$O$_{12}$ (GGG) substrate by liquid phase epitaxy (LPE). A Pt layer of 6-nm grown by dc sputtering has been used as a spin (charge) current detector (actuator) \cite{Castel2012}. The hybrid YIG/Pt system has been cut into a rectangular shape (1 mm$\times$7 mm) by using a Nd-YAG laser working at 8 W and placed on the stub line as shown in Fig. \ref{fig:Fig1} (a) with the crystallographic axis [1,1,$\bar{2}$] parallel to the planar microwave field generated by the stub line. The distance between electrodes (+ and -) is fixed at 4 mm. Figure \ref{fig:Fig1} (a) illustrates the rf design of the present study which is based on a notch filter (supplementary information in Ref. \cite{NotchVincent}) coupled to a hybrid YIG/Pt system placed at $\lambda/4$. The in-plane static magnetic field ($H$) is applied perpendicularly to the YIG/Pt bar with $H\perp$h$_{rf}$, where h$_{rf}$ corresponds to the generated microwave magnetic field. This latter configuration permits to maximize (i) the precession of the magnetization at the resonant condition and (ii) the dc voltage detection at the edges of the Pt layer. P1 and P2 illustrate the 2 ports of the vector network analyzer (VNA) that were calibrated, including both cables. The frequency range is fixed from 4 to 5 GHz at a microwave power of $P=0$ dBm. In the notch filter configuration, the maximum of magnetic field is located at $\lambda/4$ (corresponding to a short circuit, SC) and a maximum of electric field at $\lambda/2$ (open circuit, OC). Insertion of the YIG/Pt device at the SC does not impact the resonator features which is well illustrated by the frequency response of $S$ parameters (with and without the device) as shown in Fig. \ref{fig:Fig1} (b). Even if the hybrid stack includes a highly electrical conductor, no shielding effect has been observed. The $S21$ resonance peak of the empty resonator (loaded) has a half width at half maximum (HWHM) $\Delta F_{HWHM}$ of 32.75 MHz (34.45) indicating that the damping of the resonator (working at the frequency $F_{0}$) is $\beta=\Delta F_{HWHM}/F_{0}=1/2Q$=6.92$\times 10^{-3}$ (7.27$\times 10^{-3}$). This leads to a quality factor $Q$ of 72.3 (68.7). It should be noted that the latter definition of $Q$ does not reflect the electrical performance of the notch filter which is defined by $Q_{0}=F_{0}/\left[\Delta F^{S21}_{-3 dB}(1-S11_{F_{0}}) \right]$, where $\Delta F^{S21}_{-3dB}$ and $S11_{F_{0}}$ correspond respectively to the bandwidth of $S21$ at -3 dB and the magnitude (linear) of $S11$ at $F_{0}$. $Q_{0}$ is reduced from 148 to 140.5 for the empty and loaded configuration, respectively. The planar rf design used here offers many opportunities for integrated spin-based microwave applications and is not significantly affected by the hybrid YIG/Pt device. In contrast, insertion of such device (including electrical connections for the detection or actuation of the magnetization) enhanced $\beta$ by a factor of 5 \cite{Bai2015} to 12 \cite{Maier-Flaig2016}.

\section{Results and discussion}
\subsection{Strong coupling regime}

We first studied the frequency dependence of $S$ parameters at $P=0$ dBm with respect to the applied magnetic field, $H$. Figure \ref{fig:Fig1} (c) demonstrates the strong coupling regime via the anti-crossing fingerprint. The FMR and the notch filter interact by mutual microwave fields, generated by the oscillating currents in the stub and the FMR magnetization precession which led to the following features: (i) hybridization of resonances, (ii) annihilation of the resonance at $F_{0}$, and (iii) generation of two resonances at $F_{1}$ and $F_{2}$. At the resonant condition $H=H_{RES}$, the frequency gap, $F_{gap}$, between $F_{1}$ and $F_{2}$ is directly linked to the coupling strength of the system ($F_{gap}/2=g_{eff}/2\pi$). Here, the analysis is based on the harmonic coupling model \cite{Harder2016} according to which we can define the upper ($F_{1}$) and lower ($F_{2}$) branches by: 
\begin{equation}
F_{1,2}=\dfrac{1}{2}\left[ \left( F_{0}+F_{r}\right) \pm \sqrt{\left( F_{0}-F_{r}\right)^{2}+k^{4}F_{0}^{2}}\right]
\label{F12}
\end{equation}
The FMR frequency, $F_{r}$, is modelled by the Kittel equation, $F_{r}=(\gamma/2\pi)\sqrt{H(H+4\pi M_{s})}$, which is the precession frequency of the uniform mode in an in-plane magnetized ferromagnetic film. The parameter $k$ corresponds to the coupling strength which is linked to the experimental data $g_{eff}/2\pi$ by the following equation \cite{Harder2016}: $F_{gap}=F_{2}-F_{1}=k^{2}F_{0}$. A good agreement of $F_{1,2}$ (solid red and blue lines) with experimental data is obtained with $k$=0.142, the saturation magnetization $4\pi M_{s}$=1775 G, and the gyromagnetic ratio $\gamma$=1.8 10$^7$ rad Oe$^{-1}$s$^{-1}$. The color plot is associated with the $S21$ parameter for which the dark area corresponds to a magnitude of -12 dB. The same feature has been observed for negative magnetic field (not shown). Figure \ref{fig:Fig1} (d) represents the frequency dependence of the transmission spectra for the notch/YIG/Pt system for different magnetic field, lower and higher than $H_{RES}$. We find good agreement between experimental data (solid lines) and the calculated response (dash dot lines) based on the analytic solution proposed in Ref. \cite{Harder2016}, leading to an effective damping parameter of $\alpha_{eff}$=1.75 10$^{-3}$ for the YIG/Pt system. It should be noted that $\alpha_{eff}$ does not reflect the intrinsic magnetic loss but includes the inhomogeneous broadening due to the excitation of several modes. We studied the YIG/Pt sample by standard FMR using a highly sensitive wideband resonance spectrometer within a range of 4 to 20 GHz (at room temperature with a static magnetic field applied perpendicular to the sample). These characterizations lead to the intrinsic damping parameter $\alpha\approx$10$^{-4}$.

\subsection{Thermal control}
\begin{figure}
	\includegraphics[width=8cm]{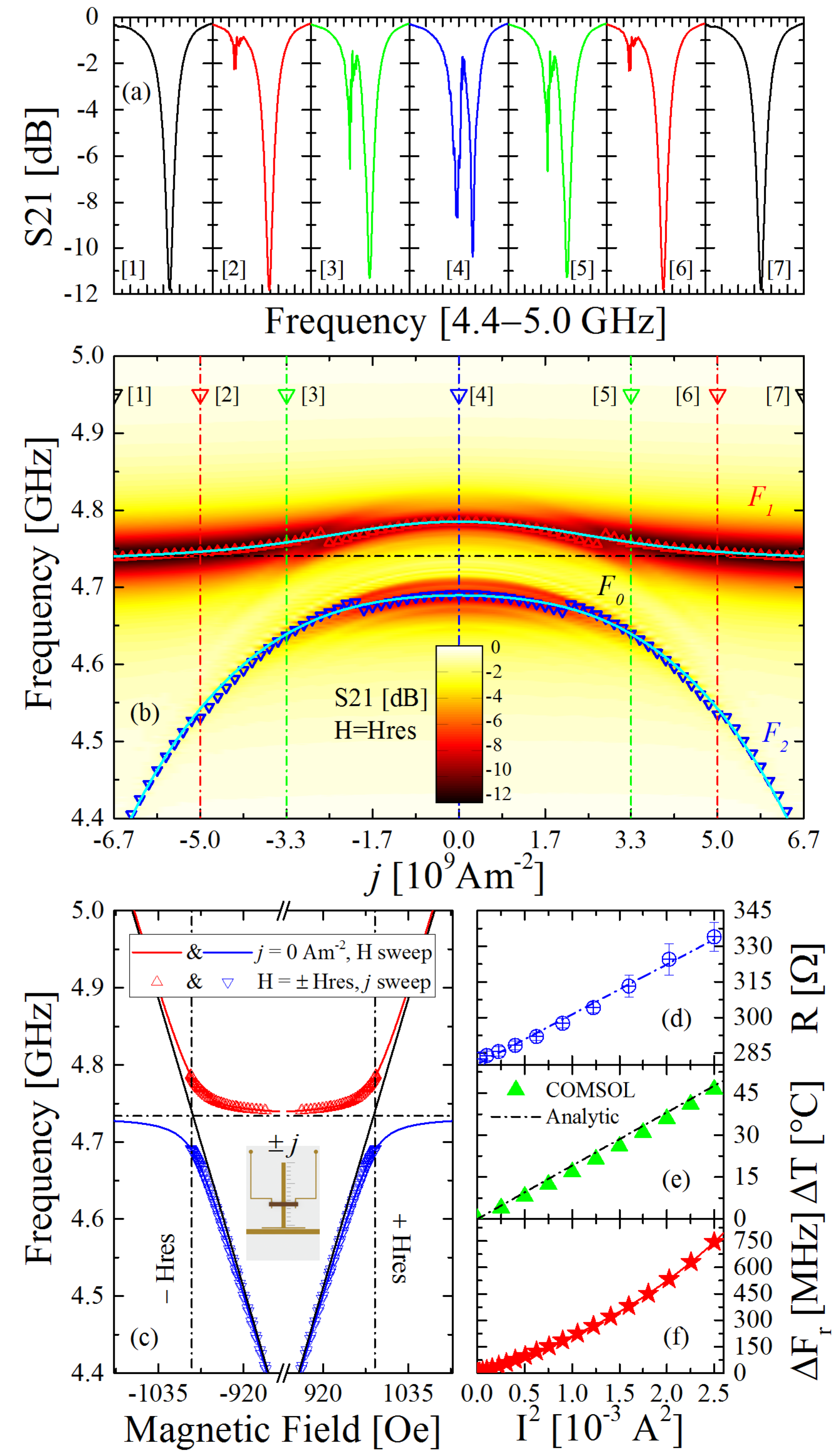}
	\caption{\label{fig:Fig4} 
		Injection: (a) Frequency dependence of the $S$ parameters at the resonant condition ($H=H_{RES}$) for different applied charge current. Labels [1] to [7] correspond to the magnitude of the applied charge current as indicated in (b). (b) Charge current dependence of the frequency measured at the resonant condition. The color map is associated to the magnitude of $S21$ from 0 to -12 dB. Solid cyan curves correspond to the current dependence of the hybridized frequencies based on Eq.(\ref{F12}), including the temperature dependence of $F_{r}$. Red and blue triangles indicate the position of $F_{1}$ and $F_{2}$. (c) Magnetic field dependence of the frequency: the correlation between magnetic field and charge current dependence of $F_{1}$ and $F_{2}$. Inset shows a schematic of the measurement setup. Dependence of Pt strip resistance (d), the temperature increase of the Pt strip (e), and $\Delta F_{r}$ (f) with respect to the injected current ($I^{2}$).
	}
\end{figure}
Sending an electric current through the structure reduces or enhances the magnetic losses, i.e. the Gilbert damping parameter $\alpha$, of the YIG film (depending on the magnetic/current setting). Furthermore, the current induced heating gives rise to a temperature difference, $\Delta T$, over the YIG/Pt system. As discussed below, the strong dependence of the YIG/Pt device to $\Delta T$ overshadows the anticipated modulation of $\alpha$ by the STT (without $\Delta T$ contribution). The absence of strong coupling in the ISHE signal (supplementary information in Appendix A) is another possible reason for the failure to control $\alpha$ of the bulk magnetization.

Figure \ref{fig:Fig4} illustrates the charge current dependence of the frequency response at the particular magnetic field, $H=H_{RES}$ at which the $S21$ splitting is maximized. Relatively small dc currents in the range of -40 to 40 mA (corresponding to a current density between -6.7 to 6.7 10$^{9}$Am$^{-2}$) were sent through the Pt contact shown in the inset from Fig. \ref{fig:Fig4} (c), which are one order of magnitude smaller than Ref. \cite{PtTemp}. The Joule heat produces a temperature gradient. Figure \ref{fig:Fig4} (a) represents the frequency dependence of the transmission spectra ($S21$ in dB) at the resonant condition ($H=H_{RES}$) for different applied charge currents. Labels [1] to [7] correspond to the magnitude of the applied charge current which is indicated in Fig. \ref{fig:Fig4} (b). The reference measurement was carried out for zero current density ([4]). By sending a current density of 3.3 ([5]), 5.0 ([6]), and 6.7 10$^{9}$Am$^{-2}$ ([7]) into the Pt strip, the resonant frequency $F_{2}$ defined by Eq.(\ref{F12}) can be significantly tuned and drastically change the transmission spectra. By reversing the sign of the charge current, the response of [1]$\&$[7] (as well as [2]$\&$[6] and [3]$\&$[5]) present the same behaviour. Figure \ref{fig:Fig4} (b) gives an overview of the latter dependence where the color map is associated to the magnitude of $S21$ from 0 to -12 dB. In this figure, we can clearly observe the relatively small changes of $F_{1}$ which approaches $F_{0}$ (resonant frequency of the notch filter) for larger current densities. This behaviour cannot be attribute to current-induced magnetic field because the response does not depend on the sign of $j$ (and the sign of $H$, as shown in Fig. \ref{fig:Fig1} (c)). Indeed, magnitude of the in-plane Oersted field can been estimated \cite{OerstedField} to not exceed $\pm$0.3 Oe (extracted in the middle of the YIG section) when $I$=$\pm$ 50 mA is sent through the Pt section of 6.10$^{-12}$m$^{2}$. Figure \ref{fig:Fig4} (c) illustrates the correlation between the effects of magnetic field and charge current on the frequency. Open blue and red triangles correspond to the experimental values of $F_{1}$ and $F_{2}$ at $H=H_{RES}$ extracted from the measurement presented in Fig. \ref{fig:Fig4} (b). A perfect equivalence of the effects of magnetic field and charge current on the resonance frequencies is evident. For example, the response of $S21$ at $H=\pm 890$ Oe (with $j=0$ Am$^{-2}$) is equivalent to $S21$ $j=\pm 6.7 10^{9}$Am$^{-2}$ (with $H=H_{RES}$) which corresponds to a shift of $\pm$100 Oe. As shown in Fig. \ref{fig:Fig4} (d), the measured resistance of the Pt layer increases quadratically with the applied current defined by $R_{(I)}=R_{0}+R_{2}I^{2}$, where $R_{2}$=21.019 k$\Omega$A$^{-2}$ and the total resistance of the Pt at room temperature is $R_{0}$=280.35 $\Omega$ corresponding to a Pt conductivity of 2.38 10$^{6}$$\Omega^{-1}$m$^{-1}$ in agreement with previous work \cite{Castel2012}.

In order to obtain quantitative information we carry out three-dimensional finite element modeling \cite{COMSOL} of our device which takes into account material and temperature dependent transport coefficients for the Pt layer \cite{PhysRevB.90.180402}. The temperature dependence of the electrical conductivity can be fitted by $\sigma (T)=\sigma /(1+a(T-T_0))$, with $\sigma$ equal to the resistance at vanishing currents, $a=4\times 10^{-3} K^{-1}$ is the temperature coefficient of resistance (TCR) of Pt which is well tabulated in the literature, and $T-T_0$ is the overall temperature increase ($\Delta T$). The thermal conductivity of the YIG substrate is kept at 5 W/m/K \cite{flipse2014observation} whereas that of Pt is calculated using the Wiedemann-Franz relation $\kappa=\sigma L_0T_0$ where $L_0$=2.44 $10^{-8}$ V$^2/$K$^2$ is the Lorenz number and $T_0$ is the reference temperature. Figure \ref{fig:Fig4} (e) presents the dependence $\Delta T$ as a function of $I^{2}$ extracted from our model (green triangles). It should be noted that the latter dependence agreements with the analytic solution \cite{PtTemp}: $\Delta T=\kappa_{Pt}(R_{(I)}-R_{0})/R_{0}$, where $\kappa_{Pt}$=254 K. We see that the temperature increase of the Pt layer is estimated at 47$\pm$0.6 K for a current density of 8.33 10$^{9}$Am$^{-2}$.

\begin{figure}
	\includegraphics[width=8.75cm]{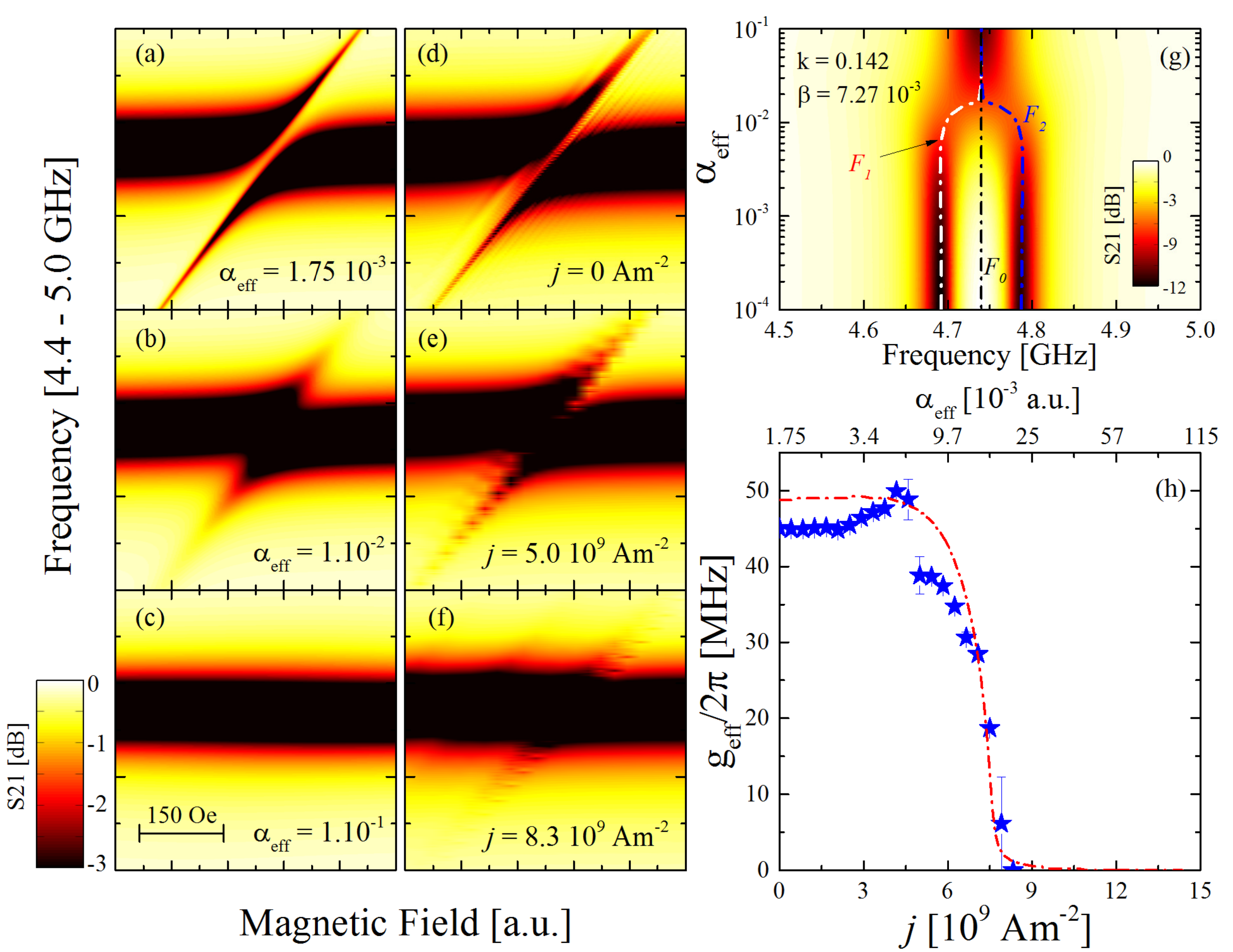}
	\caption{\label{fig:Fig5} 
		(a)-(c) Calculated magnetic field dependence of the microwave transmission spectrum based on Ref. \cite{Harder2016} for different values of the damping parameter $\alpha$. (d)-(f) Measured magnetic field dependence of the frequency as function of the injected current magnitude. The color map of (a)-(f) is associated to the magnitude of the transmission parameter $S21$ from 0 to -3 dB. (g) Calculated dependence of $S21$ (color map from 0 to -12 dB) as function of the frequency and the damping parameter. Representation of $F_{1}$, $F_{2}$, and $F_{0}$. (h) Thermal annihilation of the coupling strength measured (blue stars) and extracted from (g) (dash dot red line).		
	}
\end{figure}
Next, we measured the magnetic field dependence of the frequency response as shown in Fig. \ref{fig:Fig1} (c) as function of charge current as illustrated in Fig. \ref{fig:Fig5} (d) to (f). Figure \ref{fig:Fig4} (f) represents the dependence of $\Delta F$ as function of $I^{2}$ defined as $F_{r}(T)=F_{r}^{T_{0}=RT}-\Delta F(T)$. The observed reduction is caused by the temperature dependence of the magnetization \cite{MsT,DTMSUhcida}. Substituting this equation into Eq.(\ref{F12}) reproduces the current dependence of the hybridized frequencies, $F_{1}$ and $F_{2}$ in Fig. \ref{fig:Fig4} (c) (solid cyan curves). It should be noted that a good agreement (on $F_{1}$ and $F_{2}$) has been found as well by using the same procedure for an applied magnetic field larger than $H_{RES}$ (supplementary information in Appendix B).

A map of the coupling regime \cite{Zhang2014,Harder2016} can achieved through the representation of the dependence of $K/\alpha$ as function of $K/\beta$ where $K=k^{2}\sqrt{F_{0}/2F_{m}}$ corresponds to the effective coupling strength with $F_{m}=(\dfrac{\gamma}{2\pi})4\pi M_{s}$. Together, these latter parameters determine the coupling features of the system that can be configured from weak ($K/\alpha<$1 and $K/\beta<$1) to strong coupling regime ($K/\alpha>$1 and $k/\beta>$1). $K$ can be tuned with (i) the volume of YIG \cite{Cao2014,NotchVincent}, (ii) the volume of the cavity \cite{Schoelkopf2008}, and (iii) the magnitude of the microwave magnetic field \cite{Zhang2014,NotchVincent}. Control of the coupling regime has been already demonstrated \cite{Maier-Flaig2016,ControlBETA} by tuning the cavity losses $\beta$. However, the charge current dependence of the transmission spectra of our system suggests a control of the magnetic losses ($\alpha_{eff}$ in our case). It should be noted that our notch filter 
features remain unchanged as function of the current density sent into the Pt strip. The effect of the Pt heating is not only a shift of the resonance (tuning) but also an increased broadening (decoupling) caused by inhomogeneous heating (top part of the YIG/Pt device is hotter than the bottom part, while the microwaves see the whole sample). On the other hand, the temperature dependence of the damping parameter cannot be attributed to the Spin Seebeck Effect (SSE) because of the thickness of our sample (6 $\mu$m). Figures \ref{fig:Fig5} (a) to (c) represent the calculated magnetic field dependence of the frequency based on Ref. \cite{Harder2016} for different values of $\alpha_{eff}$ (defined above). Here we show by model calculations how to control the coupling strength by increasing the magnetic losses \cite{maier2017temperature,haidar2015thickness}. Enhancement of $\alpha_{eff}$ from 1.75 $10^{-3}$ to 1 $10^{-1}$ suppresses the frequency gap between $F_{1}$ and $F_{2}$ and thus the coupling strength. The same behaviour has been observed experimentally by changing the current density from 0 to 8.3 10$^{9}$Am$^{-2}$ as shown Fig. \ref{fig:Fig5} (d) to (f). By following the same procedure, the experimental determination of the coupling strength, $g_{eff}/2\pi$=$\dfrac{1}{2}(F_{2}-F_{1})$, at the resonant condition (by adjusting $H$ in order to keep $F_{r}=F_{0}$) has been done for several current density. The experimental (blue star) and calculated (red dash-dot line) dependences of $g_{eff}/2\pi$ are represented in Fig. \ref{fig:Fig5} (h) as function of the charge current density. Between 0 and 6 10$^{9}$Am$^{-2}$, $g_{eff}/2\pi$ is closed to 40 MHz and abruptly reduced to zero beyond this value. The calculated dependence, which permits to well reproduced the experimental behaviour, is based on the color plot from Fig. \ref{fig:Fig5} (g). It represents the dependence of the transmission spectra $S21$ (color map form 0 to -12 dB) as function of $\alpha_{eff}$ and the frequency. It should be noted that this figure represents the decoupling of the system, whereas Fig. \ref{fig:Fig4} (c) illustrates the detuning of the transmission spectra at a fixed value of the magnetic field.

\section{Conclusion}

We reported thermal control of mode hybridization between the ferromagnetic resonance and a planar resonator by using a current-induced heating method. Our set-up allows simultaneous detection of the ferromagnetic response and the dc voltage generation in the Pt layer of the system, which reveals an absence of the strong coupling signature in the ISHE signal. We demonstrate the potential for tunable filter application by electrically control the transmission spectra and the magnon-photon coupling strength at room temperature by sending a charge current through the Pt layer. 

\section*{Acknowledgments}
This work is part of the research program supported by the European Union through the European Regional Development Fund (ERDF), by Ministry of Higher Education and Research, Brittany and Rennes M\'etropole, through the CPER Project SOPHIE/STIC $\&$ Ondes, and by Grant-in-Aid for Scientific Research of the JSPS (Grant Nos. 25247056, 25220910, and 26103006).

\section*{APPENDIX A: SPIN CURRENT DETECTION}

\begin{figure}[h]
	\includegraphics[width=8cm]{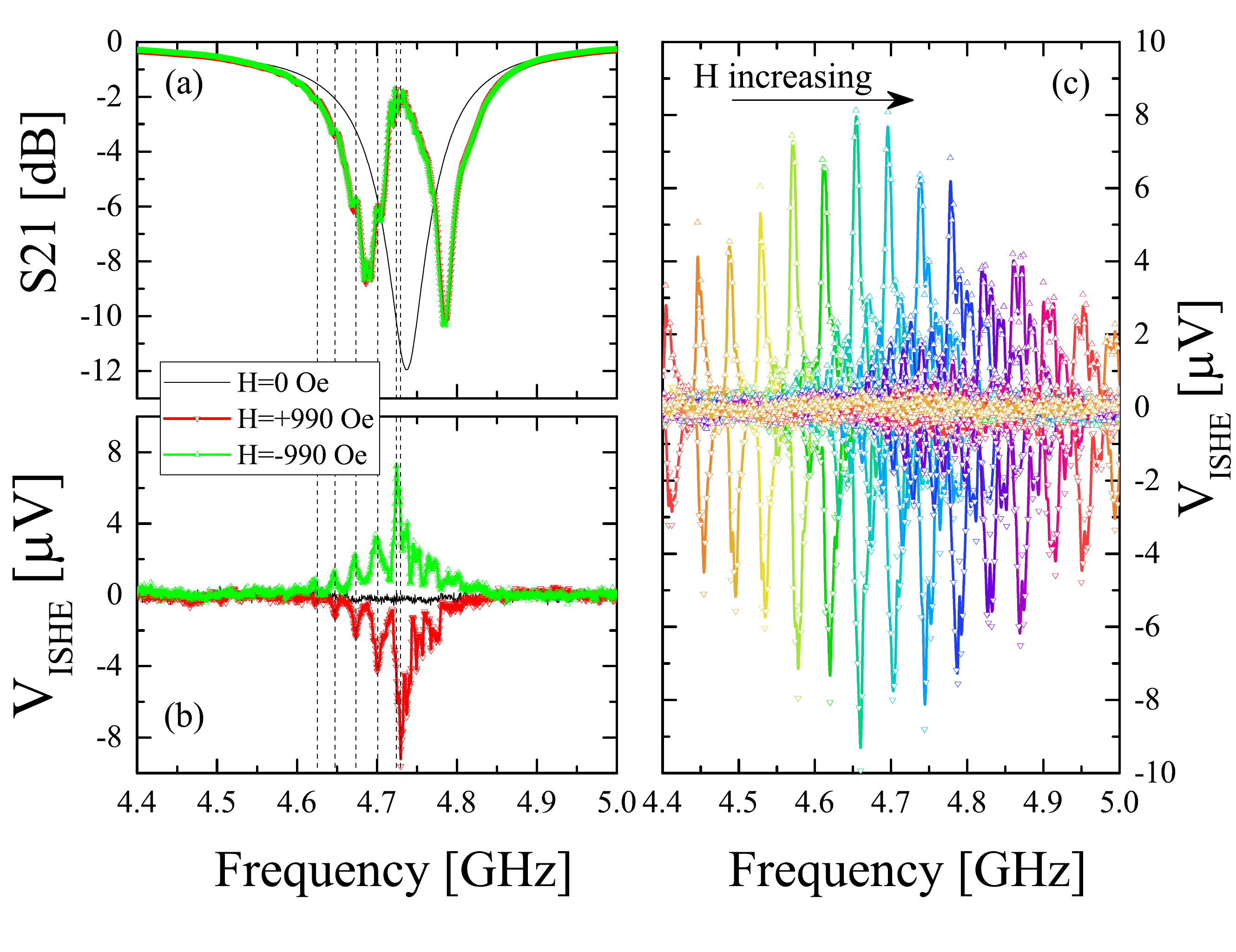}
	\caption{\label{fig:Fig3}
		Detection: Simultaneous detection of (a) $S21$ and (b) the dc voltage generated by the conversion of a spin current into a charge current via Inverse Spin Hall Effect (ISHE) at resonant condition for the positive and negative configurations of the magnetic field. (c) Measured voltage, $V_{ISHE}$, as function of the frequency for different applied magnetic field.} 
\end{figure}

Our set-up allows us to simultaneously detect the dc voltage generated in the Pt layer ($V_{ISHE}$) and the ferromagnetic response (microwave absorption) of the system as shown in Fig. \ref{fig:Fig3}. Keeping the previous calibration of the VNA, we carried out a step-by-step measurement (by shrinking the trigger mode) in order to improve the quality of the measured dc voltage by the nanovoltmeter. Figures \ref{fig:Fig3} (a) and (b) correspond respectively to the response of $S21$ and the measured voltage, $V_{ISHE}$, as function of the frequency for specific values of the applied magnetic field. A nonzero $V_{ISHE}$ comes from the fact that at (and close to) $F_{r}$ a spin current ($j_s$) is pumped into the Pt layer, which results in a dc voltage by the inverse spin Hall effect (ISHE). The sign of the electric voltage signal is changed by reversing the magnetic field and no sizable voltage is measured when $H$ is equal to zero, as expected. The sign reversal of $V$ shows that the measured signal is not produced by a possible thermoelectric effect, induced by the microwave absorption. The strong coupling in the spin pumping signal (predicted by Cao et al. \cite{Cao2014}) has been experimentally observed in a YIG(2.6-2.8$\mu$m)/Pt system for which the spin sink layer presents a thickness of 5 \cite{Maier-Flaig2016} and 10 nm \cite{Bai2015,Bai2016}. Even though the $S21$ parameter illustrates in Fig. \ref{fig:Fig1} (b) is strongly coupled, no such a signature has been observed in the dc voltage generation. We therefore conclude that in contrast to the bulk magnetization the interface sensed by the Pt layer is not strongly coupled to microwaves. However, from the comparison between Fig. \ref{fig:Fig3} (a) and (b), we are able to identify a correlation in the peak positions (vertical dashed lines) of $V_{ISHE}$ and features in $S21$. 

\section*{APPENDIX B: HYBRIDIZED FREQUENCIES}

\begin{figure}[h]
	\includegraphics[width=7.5cm]{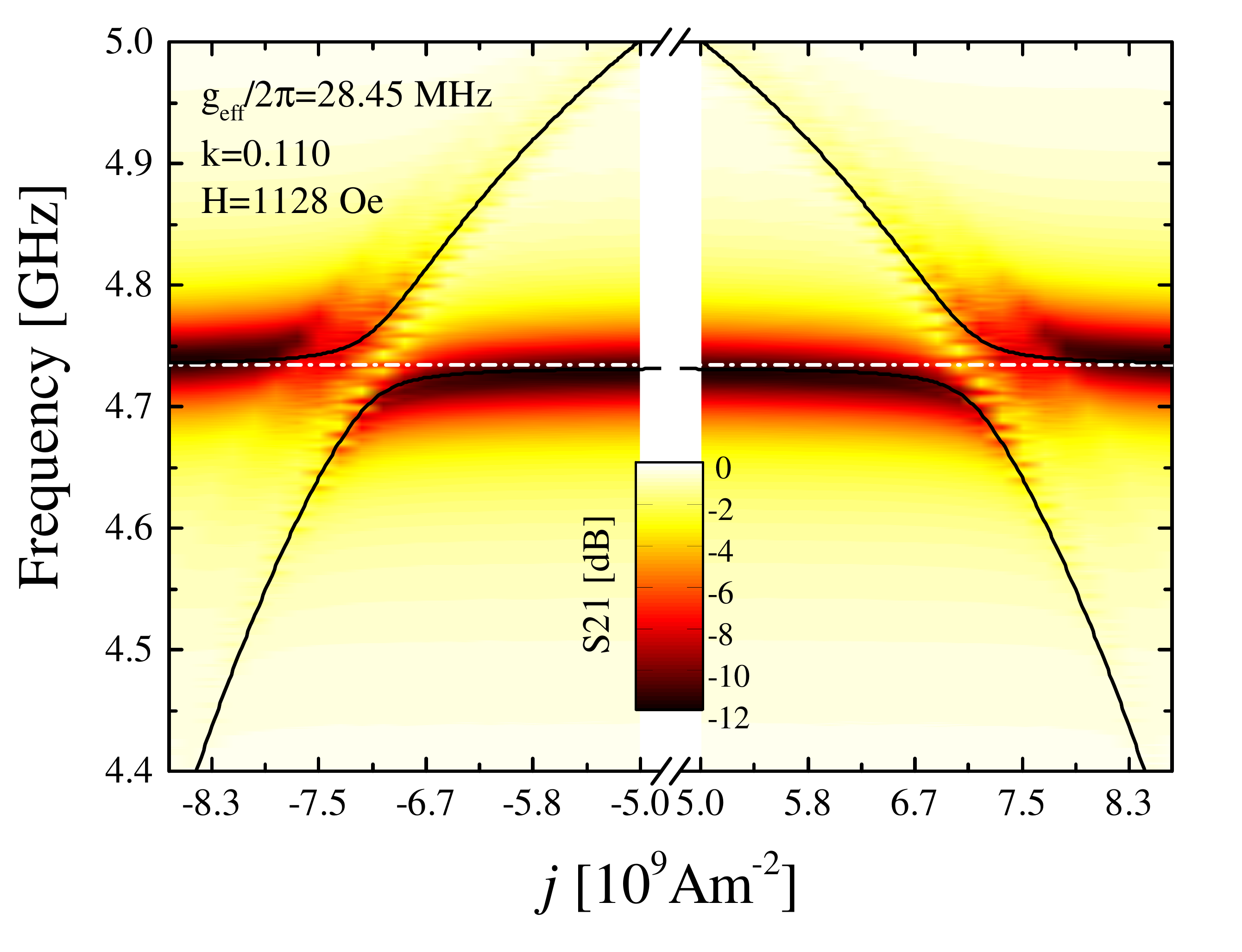}
	\caption{\label{fig:FigappB}
	Charge current dependence of the frequency measured at H=1128 Oe (off resonance). The color map is associated to the magnitude of $S21$ from 0 to -12 dB. Solid black curves correspond to the current dependence of the hybridized frequencies based on Eq.(\ref{F12}), including the temperature dependence of $F_{r}$ with $k$=0.110.
	}
\end{figure}

Figure \ref{fig:FigappB} illustrates the charge current dependence of the frequency response at an applied magnetic field where no mode hybridization is observed (H=1128 Oe). The solid black lines correspond to the calculated dependence of $F_{1}$ and $F_{2}$ based on Eq.(\ref{F12}) in which the temperature dependence of $F_{r}$ is included. It should be noted that the temperature enhancement reduces the coupling strength of our system from $k$=0.142 ($g_{eff}/2\pi\approx$45 MHz at $j=0$ Am$^{-2}$) to $k$=0.110 ($g_{eff}/2\pi\approx$28.5 MHz at $j=7.1 10^{9}$Am$^{-2}$), in agreement with current density dependence of $g_{eff}/2\pi$ represented in Fig. \ref{fig:Fig5} (g). 

\bibliography{SpinCavity}

\end{document}